\documentstyle[prl,multicol,aps,epsf]{revtex}
\begin{document}
\draft
\title{Glass-Like Heat Conduction in High-Mobility Crystalline Semiconductors} 
\author{J. L. Cohn,$^1$ G. S. Nolas,$^2$ V. Fessatidis,$^3$ T. H. Metcalf,$^4$ 
and G. A. Slack$^5$}
\address{$^1$ Department of Physics, University of Miami, Coral Gables, 
Florida 33124}
\address{$^2$ R \& D Division, Marlow Industries, Inc., 10451 Vista Park Road, 
Dallas, Texas 75238}
\address{$^3$ Department of Physics, Fordham University, Bronx, New York 10458}
\address{$^4$ Laboratory of Atomic and Solid State Physics, Cornell University, 
Ithaca, New York 14853}
\address{$^5$ Department of Physics, Rensselaer Polytechnic Institute, Troy, 
New York 12180}
\maketitle
\begin{abstract}
The thermal conductivity of polycrystalline semiconductors with type-I 
clathrate hydrate 
crystal structure is reported.  Ge clathrates (doped with Sr and/or Eu) 
exhibit lattice thermal conductivities 
typical of amorphous materials.  Remarkably, this behavior occurs in spite 
of the well-defined crystalline structure and relatively high electron 
mobility ($\sim 100$ cm$^2$/Vs).  The dynamics of dopant 
ions and their interaction with the polyhedral cages of the structure 
are a likely source of the strong phonon scattering.
\end{abstract}

\begin{multicols}{2}
\narrowtext

The low-temperature thermal conductivity ($\kappa$) of glasses has universal 
features that have been the subject of experimental and theoretical 
investigation for many years\cite{Reviews};  at the lowest temperatures 
($T<1$ K), $\kappa\propto T^2$, and for 4 K$<T<$20 K, $\kappa$ is nearly 
temperature independent (the "plateau").     
The $T^2$ behavior is attributed to phonon scattering by a broad distribution
of two-level systems, whereas the plateau arises from a strong decrease in 
the phonon mean free path with increasing frequency, e.g. due to Rayleigh 
scattering from variations in density or bond length.  Though typical of 
amorphous materials, similar 
thermal conductivity behavior, with $\kappa$ approaching theoretical 
minimum values\cite{KminSlack}, has been 
observed in some disordered crystalline dielectric 
materials\cite{KminCahill,SlackConf}.  A common characteristic 
of these latter systems is the presence of atoms or molecular groups 
having two or more semi-stable positions, and the absence of long-range 
correlation between their positions 
or orientations.  Resonant scattering of phonons by these "rattling" 
entities can also produce a plateau in $\kappa$\cite{resonances}.

The search for high-mobility semiconductors with glass-like thermal 
properties --"phonon glass, electron single crystals" (PGEC's) -- was 
initiated by Slack\cite{ThermoRevSlack}, and motivated by thermoelectric 
applications\cite{Goldsmid} which 
could be substantially improved by the higher figure of merit such 
hypothetical 
materials would possess.  It was postulated that the group IV elements 
forming clathrate-structured compounds were potential PGEC's\cite{SlackConf}. 
In this Letter we report the realization of PGEC behavior in crystalline 
Ge clathrates.  Trends in the behavior of the low-temperature 
lattice thermal conductivity for Si, Ge and Sn clathrates indicate that
the glass-like heat conductivity arises from resonant scattering of
phonons by vibrations of the dopant ions in their polyhedral cages.
The relatively high charge-carrier mobility of these
clathrates indicates that the dopant-cage interactions do not substantially 
degrade the electronic properties.

The sample preparation has been described previously\cite{ClathrateAPL}.  
The specimens 
are polycrystalline Si and Ge clathrates having the type-I hydrate crystal 
structure\cite{Structure}, doped with divalent barium (Ba$^{2+}$), strontium
(Sr$^{2+}$), and europium (Eu$^{2+}$) ions, and Sn clathrates of the same 
structure, doped with monovalent Cs$^{+}$.  The cubic crystal lattice is a
tetrahedral network of 
Si, Ge or Sn which forms periodic voids or "cages" of 20- and 24-coordinated 
polyhedra in a 1:3 ratio, respectively.  The dopant ions enter the lattice 
interstitially, residing inside these large cages.  X-ray diffraction (XRD) 
spectra from powdered clathrate samples revealed sharp lines 
with no phases other than that of type-I clathrate-hydrate ($Pm3n$).  
Trace amounts of Si were detected in the Si-clathrate sample.  
Electron beam microprobe analysis of a polished surface of each sample 
confirmed the x-ray results and yielded stoichiometric percentages for 
each sample.  In the case of the Ge-clathrates, scanning tunneling microscopy 
showed a 
defect-free 
structure within the grains.  Table I lists some of the physical and 
electronic properties 
of these phase-pure, dense polycrystalline samples.  Gallium, which 
substitutes for Si or Ge,  
was added to compensate for the charge of the divalent dopants.  
Charge neutrality in the Cs$_8$Sn$_{44}$ compound implies two Sn vacancies
per unit cell. 
A preliminary report\cite{ClathrateAPL} 
on specimens with nominal composition Sr$_8$Ga$_{16}$Ge$_{30}$ indicates 
that the carrier 
concentration and mobility vary with composition.  
Their relatively 
high mobilities (up to 100 cm$^2$/Vs) in comparison to other 
thermoelectrics\cite{MRS} suggest the 
possibility that the Ga may be ordered in these materials, an issue 
currently under investigation.

The samples were cut with a wire saw into parallelepipeds with 
approximate dimensions $5\times2\times2$ mm$^3$.  Steady-state thermal 
conductivity measurements at $T\geq4$ K were conducted in a radiation-shielded 
cryostat using a single heater and 25$\mu$m-diameter differential 
chromel-constantan thermocouple.  The electrical resistivity 
and thermoelectric power were measured simultaneously; 
room-temperature values are listed in Table I.  Heat losses 
due to radiation and conduction through the leads were measured 
in separate experiments and the data corrected.  These losses 
were 10-15\% near room temperature and $<$ 5\% for $T<$120 K.  The 
absolute accuracy in $\kappa$ is $\pm10$\%, limited by uncertainty in the 
thermocouple junction separation.  Measurements at $T<4$ K were conducted
in a dilution refrigerator using a similar technique with two thermometers.

Figure 1 shows lattice thermal conductivities, $\kappa_g$=$\kappa-\kappa_e$,
in the range 5 K$<T<$300 K for the five clathrates listed in Table I. The 
electronic thermal conductivity, $\kappa_e$, was estimated using the 
electrical resistivities and Wiedemann-Franz law with ideal 
Lorenz number, $L_0=2.44\times10^{-8}$ W$\Omega$/K$^2$.  Also shown is 
$\kappa$ for amorphous 
germanium (a-Ge) and a-SiO$_2$\cite{KaGe}.  
The $T$=295 K values of $\kappa_g$
for all of the specimens are small, lying between that of a-SiO$_2$ and a-Ge, 
and approaching $\kappa_{min}$ for Ge, the minimum theoretical value calculated 
using the model of Slack\cite{KminSlack}, and taking the minimum mean free 
path (mfp) of 
the acoustic phonons as one-half their wavelength\cite{KminCahill}.  
The Sn clathrate exhibits a temperature dependence typical of crystalline 
insulators, with $\kappa_g$ increasing with decreasing $T$ approximately 
as $1/T$, the signature of
propagating phonons scattered by anharmonic interactions. 
For the Si-clathrate, $\kappa_g$ also increases with decreasing $T$,
but not as strongly, indicating additional scattering at low temperatures.
The Ge clathrates have $\kappa_g$ suppressed below that of the Sn clathrate 
by more than an order of magnitude at low $T$.
Most striking are the plateaus or minima in $\kappa_g(T)$ for the Ge clathrates, 
similar 
to that of a-SiO$_2$ and other amorphous materials in the range 4 K$<T<$30 K.  
This similarity extends to $T<1$ K as shown 
in Fig.~2 for the Sr$_8$Ga$_{16}$Ge$_{30}$ specimen; $\kappa_g\propto T^2$ 
is observed down to $T$=60 mK. Our previous estimates of grain-boundary 
scattering for the Ge clathrates\cite{ClathrateAPL} and preliminary $\kappa$ 
measurements on Ge clathrates with similar 
composition and mm-sized grains\cite{MRS}, indicate that grain boundaries are 
not the source of the glass-like $\kappa$ in these materials. 

To interpret the $\kappa_g$ data in the absence of experimental 
studies of the vibrational spectra for these
semiconducting clathrates, it is instructive 
to consider prior work on the thermal conductivity\cite{Khydrates} 
and vibrational properties\cite{HydrateNeutrons}
of type I clathrate hydrates (ice clathrates).  These
materials consist of 
similar cage-like structures,\break
\centerline{\epsfxsize=3.25in\epsfbox[50 60 540 750]{clathf1.psc}}
\vskip -1.68in
\begin{figure}
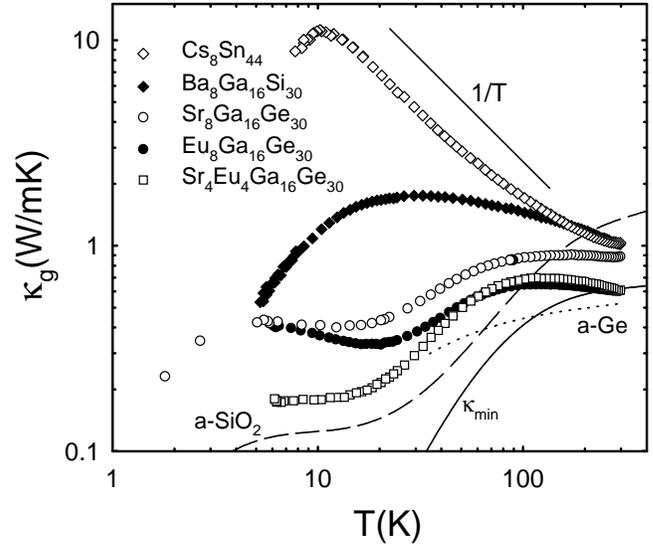

\caption{Lattice thermal conductivity vs. temperature for 
the five clathrate specimens listed in Table~I.
The dashed and dotted curves are for a-SiO$_2$ and a-Ge 
(Ref.~\protect{\onlinecite{KaGe}}), and the solid curve is 
the calculated minimum thermal conductivity of Ge (see text).}
\label{KvT}
\end{figure}
\noindent
comprised of water molecules,
which encapsulate gas atoms or molecules (guests).  Their thermal 
conductivities 
are anomalously low and glass like. The qualitative features of the 
guest-host
interactions for these materials are expected to be similar to 
those of semiconducting clathrates.
Neutron scattering studies and 
molecular/lattice dynamical calculations\cite{MDcalcs} have established 
that localized guest vibrations interact strongly with the host acoustic 
phonon branches in the clathrate hydrates.
This interaction arises as an avoided crossing of dispersing 
phonon branches and non-dispersing guest vibrations of the same symmetry,
and yields a resonant damping of heat-carrying phonons at certain points 
in the Brillouin zone.  
The guest 
translational vibration
frequencies increase as the size difference between guest and host cage 
decreases, a natural consequence of stronger restoring forces for the guests.
Generally, one expects two prominant resonances, with the lower (higher) 
frequency corresponding to guest vibrations in the larger (smaller) host 
cages.
\end{multicols}
\widetext
\begin{table}
\caption{The five compounds of this study with atomic percentages from 
electron-beam microprobe analysis, lattice parameters, $a_0$, average grain sizes, 
room-temperature electrical resistivities, $\rho$, and absolute 
Seebeck coefficients, $S$.
\label{table1}}
\begin{tabular}{cccddd}
compound&Elemental atomic \%&$a_0$(\AA)&grain size ($\mu$m)&
$\rho$(m$\Omega$cm)& S($\mu$V/K)\\
\tableline
Cs$_8$Sn$_{44}$&15.6Cs/84.4Sn&12.113&13.0&59.4&-304\\
Ba$_8$Ga$_{16}$Si$_{30}$&15.5Ba/27.5Ga/57.0Si&10.554&9.6&2.04&-66\\
Sr$_8$Ga$_{16}$Ge$_{30}$&14.8Sr/30.2Ga/55.3Ge&10.731&9.6&12.8&-313\\
Eu$_8$Ga$_{16}$Ge$_{30}$&14.2Eu/28.3Ga/57.6Ge&10.711&5.4&2.52&-152\\
Sr$_4$Eu$_4$Ga$_{16}$Ge$_{30}$&8.2Sr/6.9Eu/28.6Ga/56.4Ge&10.726&12.2&1.00&-88\\
\end{tabular}
\end{table}
\widetext
\begin{multicols}{2}
\centerline{\epsfxsize=3.25in\epsfbox[50 60 540 770]{clathf2.psc}}
\vskip -2in
\begin{figure}
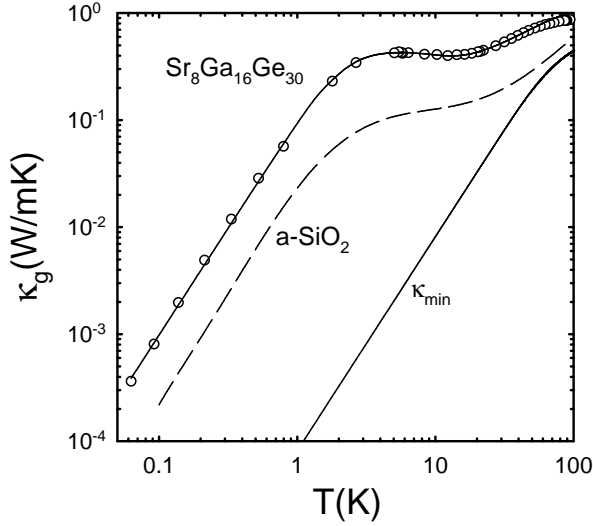

\narrowtext
\caption{Lattice thermal conductiviy measured from 60 mK to 100 K 
for Sr$_8$Ga$_{16}$Ge$_{30}$ with a fit (solid curve) to the model discussed in
the text (see also Fig.~3).  The dashed line is for a-SiO$_2$, 
and the solid curve labelled $\kappa_{min}$ is the calculated minimum thermal 
conductivity of Ge.}
\label{LowT}
\end{figure}
Consider the relevant guest and host dimensions for our specimens.  The 
radii of the Ba$^{2+}$, Eu$^{2+}$ and Sr$^{2+}$ ions are 2.07, 
1.91 and 1.90 \AA, 
respectively, from x-ray crystallographic data of the respective 
telluride compounds with the NaCl-type structure\cite{Structure}.  These 
dimensions may not be accurate for the Si or Ge-clathrates where the ions 
reside in large polyhedral ``cages'', but the trend in radii should be similar.
The Si$_{20}$ and Si$_{24}$ cage radii are estimated to be 2.05 and 2.23 \AA, 
respectively, from XRD data on Si-clathrates doped with alkali 
metals\cite{Cageradii}.  XRD also indicates no significant influence of Ga 
substitution on the average size of the cages.  The Ge$_{20}$ 
and Ge$_{24}$ cages have radii of 2.15 and 2.32 {\rm\AA}, 
respectively\cite{Cageradii}.  Thus we expect that Eu$^{2+}$ and Sr$^{2+}$ in 
the Ge compounds are ``looser'' in their cages than are Ba$^{2+}$ ions in 
the Si compound.  We infer that the ionic radius for Cs$^+$ in the 
Sn clathrate is approximately
the same as the cage size from the following analysis.  Using data for 
isostructural K- and Rb-doped Sn clathrates from the 
literature\cite{Snclathrates} 
we plot lattice constant {\it vs} alkali-metal ion radius\cite{Shannon}, 
and observe that the measured lattice constant for 
Cs$_8$Sn$_{44}$ is larger than that implied by linear 
extrapolation on this plot.  This suggests that Cs$^+$
in the cages of Cs$_8$Sn$_{44}$ causes a lattice expansion.

From these observations it is apparent that the magnitude of 
the $T=10$ K lattice thermal resisitivity, $1/\kappa_g$, correlates 
with the relative guest/host-cage size mismatch.  This motivates our proposal 
that
the plateaus or dips in $\kappa_g$ for the Ge clathrates are associated with 
resonant scattering via dopant vibrations.  The absence of this 
behavior in the Sn clathrate and weaker effect in the Si clathrate 
suggest that the vibrational\break
\centerline{\epsfxsize=3.25in\epsfbox[50 60 540 770]{clathf3.psc}}
\vskip -1.7in
\begin{figure}
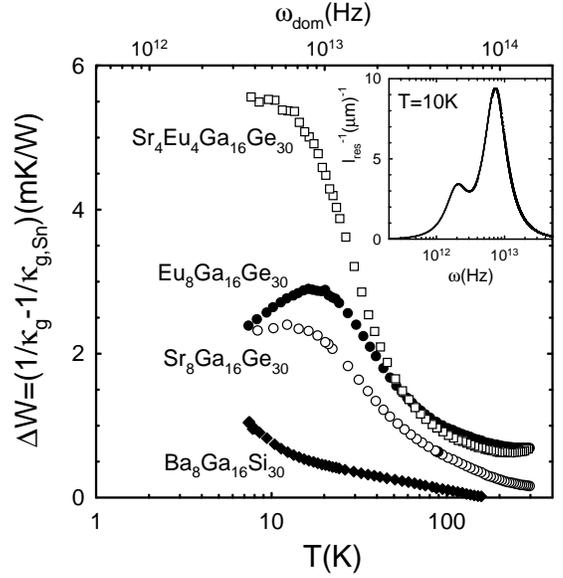

\narrowtext
\caption{Difference in the lattice thermal resistivity of each Ge clathrate
from that of the Sn clathrate.  The frequency scale (upper ordinate) is for
dominant thermal phonons ($\hbar\omega=3.8k_BT$).  The inset shows the 
inverse scattering length at $T=10$K for the resonant terms (see text) used 
to fit the data for Sr$_8$Ga$_{16}$Ge$_{30}$ (solid curve in Fig.~2). The 
parameters were: $C_1=3.8\times10^{6}$m$^{-1}$s$^{-3}$K$^{-2}$, 
$C_2=1.7\times 10^{8}$m$^{-1}$s$^{-3}$K$^{-2}$, 
$\omega_1=1.7$THz, $\omega_2=6.3$THz, and $\gamma_1=\gamma_2=0.8$.}
\label{DeltaW}
\end{figure}
\noindent
frequencies of Cs$^{+}$ and Ba$^{2+}$, 
respectively, are sufficiently high in energy that a strong interaction with 
the acoustic phonon branches of the host does not occur for these compounds. 
To further explore the resonance scattering within this scenario we use 
the Sn-clathrate data as a reference, 
computing the difference in thermal resistivity for each of the other 
clathrates
from that of the Sn clathrate, $\Delta W=1/\kappa_g-1/\kappa_{g,Sn}$.
These results are plotted in Fig.~\ref{DeltaW} and, if this interpretation is 
correct, represent a measure of the thermal resistivity due to resonance 
scattering in the Si and Ge clathrates.  We see that $\Delta W$ for the Ge 
clathrates is peaked in 
the temperature range 10-20 K.  Within the dominant phonon 
approximation ($\hbar\omega_{dom}\simeq 3.8k_BT$ for constant mfp) 
this corresponds
to phonon frequencies $\omega_{dom}\sim 5-10$THz (upper ordinate in Fig.~3).  
The asymmetry of the maxima for the Sr- and Eu-doped specimens suggests 
the presence of a second, smaller resonance peak at lower frequency.

The $\kappa_g\propto T^2$ behavior at low temperatures implies a phonon mfp 
inversely proportional to frequency, $l(\omega)\propto\omega^{-1}$.
This follows from the kinetic theory expression, $\kappa_g=Cvl/3$ ($C$ and $v$ 
are the lattice specific heat and sound velocity, respectively), 
a Debye specific heat, $C\propto T^3$, and $\omega\propto T$ for thermal 
phonons.   
The localized tunnel systems (TS) that give rise to this
phonon scattering in amorphous materials\cite{Reviews} also yield anomalous 
features in the specific heat and elastic properties in the same temperature 
range.  It is
plausible that TS's are the source of the scattering in the clathrates.  
For example, they could correspond to different 
positions of the dopant ions in their cages; a broadened density
of these states might be induced by the random distribution of the
dopants.  Measurements of specific heat and elastic properties 
are needed to address this issue.

At $T\le 40$ K, where $\kappa_g\gg\kappa_{min}$ and phonon transport is an 
appropriate description, the behavior of $\kappa_g$ for the Ge specimens 
can be simulated by 
integrating the kinetic theory expression, 
$\kappa_g=(v/3)\int{C(\omega)l(\omega)}$, with a Debye
specific heat and a mfp that is 
a sum of terms representing TS, resonant, and Rayleigh (R)
scattering, $l(\omega)=(l_{TS}^{-1}+l_{res}^{-1}+l_R^{-1})^{-1}+l_{min}$, 
with\cite{TSPhillips,TSAnderson,Graebner},
\begin{eqnarray*}
l_{TS }^{-1}=&&A(\hbar\omega/k_B){\rm tanh}(\hbar\omega/2k_BT) \\
&&+(A/2)(k_B/\hbar\omega+B^{-1}T^{-3})^{-1}, \\
l_{res}^{-1}=&&\sum_i C_i\omega^2T^2/[(\omega_i^2-\omega^2)^2+
\gamma_i\omega_i^2\omega^2],\\
l_R^{-1}=&&D\hbar^4\omega^4/k_B^4. \nonumber
\end{eqnarray*}
The constants $A$ and $B$ are related to microscopic 
variables describing the TS model\cite{Graebner}.  The phenomenological 
resonance terms are of the form employed previously to describe phonon 
scattering in ionic crystals\cite{Pohl}.  The lower limit on $l$ 
is assumed to be a constant, $l_{min}$.   Using $v=3370$ m/s and 
$\Theta_D=360$ K
appropriate for diamond-structured Ge\cite{GeDebye}, the solid curve in Fig.~2 
was generated for Sr$_8$Ga$_{16}$Ge$_{30}$
using $A=1.38\times 10^4$ m$^{-1}$K$^{-1}$, $B=1.5\times10^{-3}$ K$^{-2}$, 
$l_{min}=3$ {\rm\AA}, $D=1$ m$^{-1}$K$^{-4}$ and a sum of two resonance 
terms shown in the inset of Fig.~3.  The TS parameters are
comparable to those found for many amorphous solids\cite{Graebner}.  
The $l_{res}^{-1}$ employed is not unique and the values of $\omega_1$ and 
$\omega_2$ are sensitive to the value of $D$ in the range characteristic of 
crystalline materials, $0\leq D\leq 5$ m$^{-1}$K$^{-4}$ 
(two orders of magnitude smaller than 
in glasses\cite{Graebner}). Within this range of $D$, we achieved good fits 
for resonance frequencies, 0.6~THz$\leq\omega_1\leq 2$ THz and 
3 THz$\leq\omega_2\leq 20$ THz.
The data for the other Ge specimens
could be fitted using the same TS parameters by simply adjusting the 
magnitudes of the two resonance terms without substantially altering
the resonance frequencies. 
Thus a sum of two resonances has the minimum complexity required to
reproduce the range of $\kappa_g$ behavior we have observed.  We expect a 
``freeze-out'' of the guest-ion rattling at low temperatures; this should 
give rise to static disorder in bond lengths
associated with a distribution of the guest-ion positions within their cages.  
Small energies ($\leq 1$ K) presumably characterize the separation between 
these states, and are the origin of the TS.

These unique materials possess interesting and potentially useful properties
that call for further investigation.  As one of only a few crystalline 
materials having glass-like thermal conduction, other thermal and elastic 
properties at $T<1$ K are of particular interest.

\end{multicols}
\end{document}